\documentclass[twocolumn,aps,prl,superscriptaddress,10pt]{revtex4}
\usepackage{amsmath}
\usepackage{amsfonts}
\usepackage{amssymb}
\usepackage{mathrsfs}
\usepackage{graphicx}
\usepackage{epstopdf}
\usepackage{subfigure}
\usepackage[colorlinks,
            linkcolor=blue,
            anchorcolor=blue,
            citecolor=blue,
            urlcolor=blue
            ]{hyperref}
\usepackage{txfonts}%

\providecommand{\U}[1]{\protect\rule{.1in}{.1in}}

\begin{document}
\title{Tunable non-Hermitian skin effect via gain and loss}
\author{Wen-Cheng Jiang}\thanks{These two authors contributed equally.}
\affiliation{School of Science, Chongqing University of Posts and Telecommunications, Chongqing 400065, China}

\author{Hong Wu}\thanks{These two authors contributed equally.}
\affiliation{School of Science, Chongqing University of Posts and Telecommunications, Chongqing 400065, China}

\author{Qing-Xu Li}
\affiliation{School of Science, Chongqing University of Posts and Telecommunications, Chongqing 400065, China}
\affiliation{Institute for Advanced Sciences, Chongqing University of Posts and Telecommunications, Chongqing 400065, China}
\affiliation{Southwest Center for Theoretical Physics, Chongqing University, Chongqing 401331, China}

\author{Jian Li}
\email{jianli@cqupt.edu.cn}
\affiliation{School of Science, Chongqing University of Posts and Telecommunications, Chongqing 400065, China}
\affiliation{Institute for Advanced Sciences, Chongqing University of Posts and Telecommunications, Chongqing 400065, China}
\affiliation{Southwest Center for Theoretical Physics, Chongqing University, Chongqing 401331, China}

\author{Jia-Ji Zhu}
\email{zhujj@cqupt.edu.cn}
\affiliation{School of Science, Chongqing University of Posts and Telecommunications, Chongqing 400065, China}
\affiliation{Institute for Advanced Sciences, Chongqing University of Posts and Telecommunications, Chongqing 400065, China}
\affiliation{Southwest Center for Theoretical Physics, Chongqing University, Chongqing 401331, China}
\date{\today}

\begin{abstract}
We investigate theoretically tunable non-Hermitian skin effect in systems with gain and loss, and find that bipolar (quadripolar) non-Hermitian skin effect characterized by topological invariants in one (two)-dimensional system. We also find the partial non-Hermitian skin effect with the coexistence of localized states and extended states. Both types of the non-Hermitian skin effect have not yet been predicted together in a single system. A feasible experimental scheme of our model is proposed to realize in electric circuits. Our investigation unveils a new type of non-Hermitian skin effect and enhance the tunability of the non-Hermitian systems by gain and loss other than the conventional non-reciprocal hopping.
\end{abstract}
\maketitle
\date{today}

\textit{Introduction}.--- It is well known that the Hamiltonian is assumed to be Hermitian in conventional quantum mechanics. However, most realistic systems contain dissipation effects, and inevitably introduce non-Hermitian Hamiltonians\cite{Ashida2020,RevModPhys.93.015005,Ding2022,YU20241}. In recent years, the studies of various dissipation systems, such as in optical \cite{PhysRevLett.115.040402,Han2019,PhysRevLett.123.165701,PhysRevResearch.3.023211,Xia2021,Dai2024}, phononic \cite{PhysRevResearch.2.013280,PhysRevLett.129.084301,Zhou2023}, acoustic \cite{Hu2021,Zhang2021,Gu2022,PhysRevLett.131.066601,WAN20232330,Huang2024}, electric circuit systems\cite{Choi2018,PhysRevB.99.201411,PhysRevApplied.13.014047,Helbig2020,Zou2021,PhysRevLett.126.215302,PhysRevB.105.195127,PhysRevLett.130.057201,PhysRevB.107.085426}, and some metamaterials \cite{liu2023,RevModPhys.96.015002}, have revealed rich non-Hermitian physics. The unique features of non-Hermitian systems include the non-Hermitian skin effect (NHSE), a large number of bulk states localized at the edges \cite{PhysRevLett.121.086803,PhysRevLett.123.246801,Li2020,PhysRevLett.125.186802,PhysRevLett.127.116801,zhang2022,PhysRevA.106.053315,PhysRevB.108.L220301,jiang,e25101401,Lin2023}, and exceptional points where both the energy eigenvalues and corresponding eigenstates coalesce \cite{PhysRevLett.80.5243,PhysRevLett.116.133903,Feng2017,PhysRevLett.121.026808,PhysRevA.98.052116,PhysRevX.9.041015,PhysRevA.102.023308,PhysRevA.102.053510,PhysRevB.103.235110,PhysRevB.108.085134,PhysRevLett.132.220402}. These novel effects hold significant promise for potential applications, such as non-Hermitian topological sensor with high sensitivity \cite{Lau2018,Hokmabadi2019,science,Soleymani2022,Yang2023,Li2023,hu2024synthetically}, topological laser \cite{Longhi2018,Harari2018,PhysRevLett.121.073901,PhysRevLett.125.033603,PhysRevLett.129.013903,PhysRevLett.131.023202,Leefmans2024} and enhanced energy harvesting \cite{CommPHYSICS2021}.

The NHSE is gaining significant attention for its resilience against disturbances through the topological property characterized by the spectral winding number and its wide range of applications. It can be well described by the theory of generalized Brillouin zone (GBZ) \cite{PhysRevLett.121.086803,PhysRevX.14.021011}. For the localized states, the traditional Bloch wave vector is replaced by $k \to k'=k-i\ln |\beta|$, where the $\beta=e^{ik'}$ is the so-called GBZ. The non-Bloch band theory with GBZ introduces the non-Bloch bulk-boundary correspondence of non-Hermitian topological insulators \cite{PhysRevB.99.081103,PhysRevLett.123.066404,PhysRevLett.125.226402,PhysRevLett.125.126402,PhysRevB.102.041119,PhysRevB.103.L041404}. The presence of higher dimensions, additional symmetries, or more intricate systems will lead to richer phenomena of NHSE, such as the higher-order NHSE\cite{PhysRevB.102.205118,PhysRevB.102.241202,PhysRevB.103.045420,PhysRevB.103.L041115,Palacios2021,PhysRevLett.131.116601}, hybrid skin-topological effect \cite{PhysRevLett.123.016805,PhysRevLett.124.250402,PhysRevB.106.035425,PhysRevLett.128.223903,PhysRevB.107.195112,PhysRevB.108.075122,Chen_2024,PhysRevLett.132.063804}, and geometry-dependent NHSE \cite{YZS2022,PhysRevLett.131.207201,Zhou2023,Huang2024}, chiral NHSE \cite{PhysRevResearch.6.013213,PhysRevLett.132.113802}, and bipolar NHSE \cite{PhysRevLett.123.246801,PhysRevB.109.045410,jiang}.

\begin{figure}[hbth]
\includegraphics[width=0.47\textwidth]{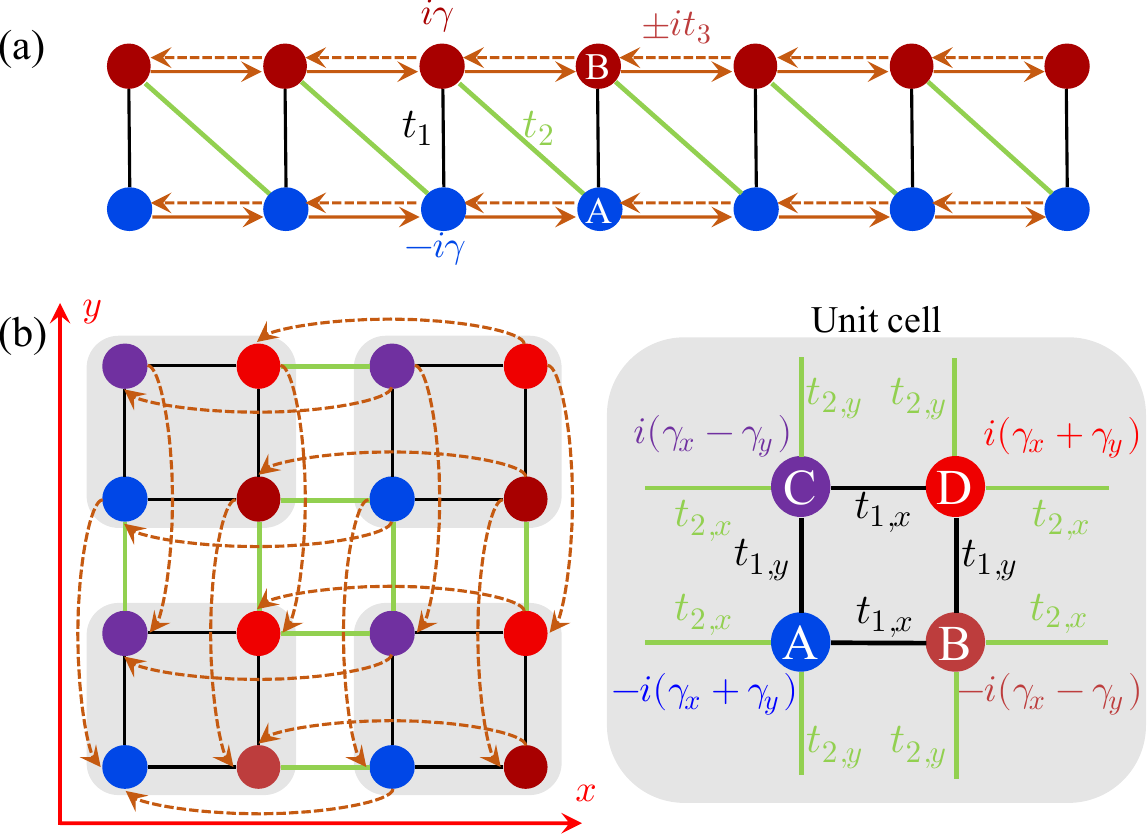}
\caption{Schematics of our model. (a) The 1D Su-Schrieffer-Heeger (SSH) ladder. This system consists of two sublattices A and B, indicated by crimson and blue circles, respectively. The parameters $t_{1}$ and $t_{2}$ are the intracell and intercell hopping rates, respectively. $\pm it_3$ are the hopping between sites in a sub-chain. The terms $\pm i\gamma$ represent the on-site gain and loss. (b) Schematic of a 2D SSH model on a square lattice.  There are four sublattices (A, B, C, and D) in one unit cell. The intra-cell and inter-cell hopping amplitudes are $t_1$ and $t_2$, respectively. The orange dotted lines correspond to long-range hoppings between unit cells with strength $\pm it_3$, and arrows represent $it_3$.
}%
\label{fig:1}%
\end{figure}

However, the tunability of NHSE is highly demanded for thorough comprehension and practical applications of non-Hermitian systems. Previous studies show that asymmetric coupling and/or gain/loss (complex hopping) can lead to NHSE \cite{PhysRevLett.116.133903,PhysRevLett.125.186802,PhysRevLett.128.223903,PhysRevB.104.155412,PhysRevLett.128.120401,PhysRevB.106.085427,PhysRevA.106.062206,PhysRevB.109.045410,LIAO2024107372}. Yet, the tunable asymmetric hopping can hardly be implemented in usual non-Hermitian systems due to the absence of inherent non-reciprocity. One might wonder how it is possible to fine-tune the NHSE exclusively through gain and loss, without needing non-reciprocity.

In this work, we investigate the NHSE in both one and two-dimensional system with gain and loss. The one-dimensional (1D) system experiences phase transitions from the $\mathcal{PT}$-symmetric phase through $\mathcal{PT}$-broken phase I to $\mathcal{PT}$-broken phase II with consistently growing gain and loss. There are partial NHSE and bipolar NHSE in the two $\mathcal{PT}$-broken phase respectively. The partial NHSE shows coexistence of localized state and extended states. For the two-dimensional (2D) system, we can find a novel quadripolar NHSE instead of the bipolar NHSE with localized states at the four corners of the systems. Finally we propose a experimental feasible scheme to realize our model in a practical system of electric circuits.

\textit{Dispersions and NHSEs in 1D non-Hermitian system}.---We consider a non-Hermitian fermionic system on 1D ladder with balanced gain and loss [see FIG. \ref{fig:1}(a)]. The corresponding Hamiltonian reads
\begin{eqnarray}
\begin{aligned}
H&=\sum_{n=1}^{N}t_{1}\left( c_{n,a}^{\dagger} c_{n,b}+ h.c.\right)
+\sum_{n=1}^{N-1} \Big (t_{2} c_{n,b}^{\dagger} c_{n+1,a}\\
&+t_{3}e^{i\phi} c_{n,a}^{\dagger} c_{n+1,a}+t_{3} e^{i\phi} c_{n,b}^{\dagger} c_{n+1, b}+h.c.\Big )\\
&+\sum_{n=1}^{N}\left(-i \gamma c_{n,a}^{\dagger} c_{n,a}+i \gamma c_{n,b}^{\dagger} c_{n,b}\right).
\label{eq:1}
\end{aligned}
\end{eqnarray}
where $c_{n,\alpha}$ ($c_{n,\alpha}^\dagger$) represents the annihilation (creation) operator of spinless fermions on $\alpha$ ($\alpha=a$ or $b$) sublattice of $n$th unit-cell. The parameters $t_{1}$ is the intracell hopping, $t_2$ is the intercell hopping between different sublattices, and $\pm it_3$ is the intercell hopping between the same sublattice. The terms $i\gamma$ ($-i\gamma$) represents the on-site gain (loss). Both $\gamma$ and $t_{1,2,3}$ are real parameters, and $N$ is the number of unit cells. Under periodic boundary conditions (PBCs), we perform the Fourier transformation of Hamiltonian Eq.(\ref{eq:1}) and arrive at $H=\sum_k\Psi^{\dag}_k h(k) \Psi_k$ with $\Psi_k=[c_{a,k},c_{b,k}]^{T}$, where $c_{a,k}$ ($c_{b,k}$) is the Fourier transformation of $c_{n,a}$ ($c_{n,b}$). The corresponding Bloch Hamiltonian is given by
\begin{equation}
\begin{aligned}
h(k)  &= \left(t_{1}+t_{2} \cos k\right) \sigma_{x}+t_{2} \sin k \sigma_{y} \\
&-i \gamma \sigma_{z}-2t_{3}\sin(k) \mathbb{I},
\label{eq:2}
\end{aligned}
\end{equation}
where $\sigma_{x,y,z}$ and $\mathbb{I}$ represent the Pauli matrices and the identity matrix, respectively.
%In the case ($\gamma=\phi=0$), the Hamiltonian possesses both inversion $\mathcal{P}=\sigma_x$ and time-reversal $\mathcal{T}=\mathcal{K}$ symmetries.
The non-vanishing $\gamma$ and $\pm it_3$ can break both $\mathcal{P}$ and $\mathcal{T}$ symmetry, the system still preserves parity-time ($\mathcal{PT}$) symmetry. Besides, the system has anomalous particle-hole symmetry due to
\begin{equation}
\sigma_z    h^*(-k)  \sigma_z=-h(k).
\end{equation}

\begin{figure}[hbth]
\includegraphics[width=0.5\textwidth]{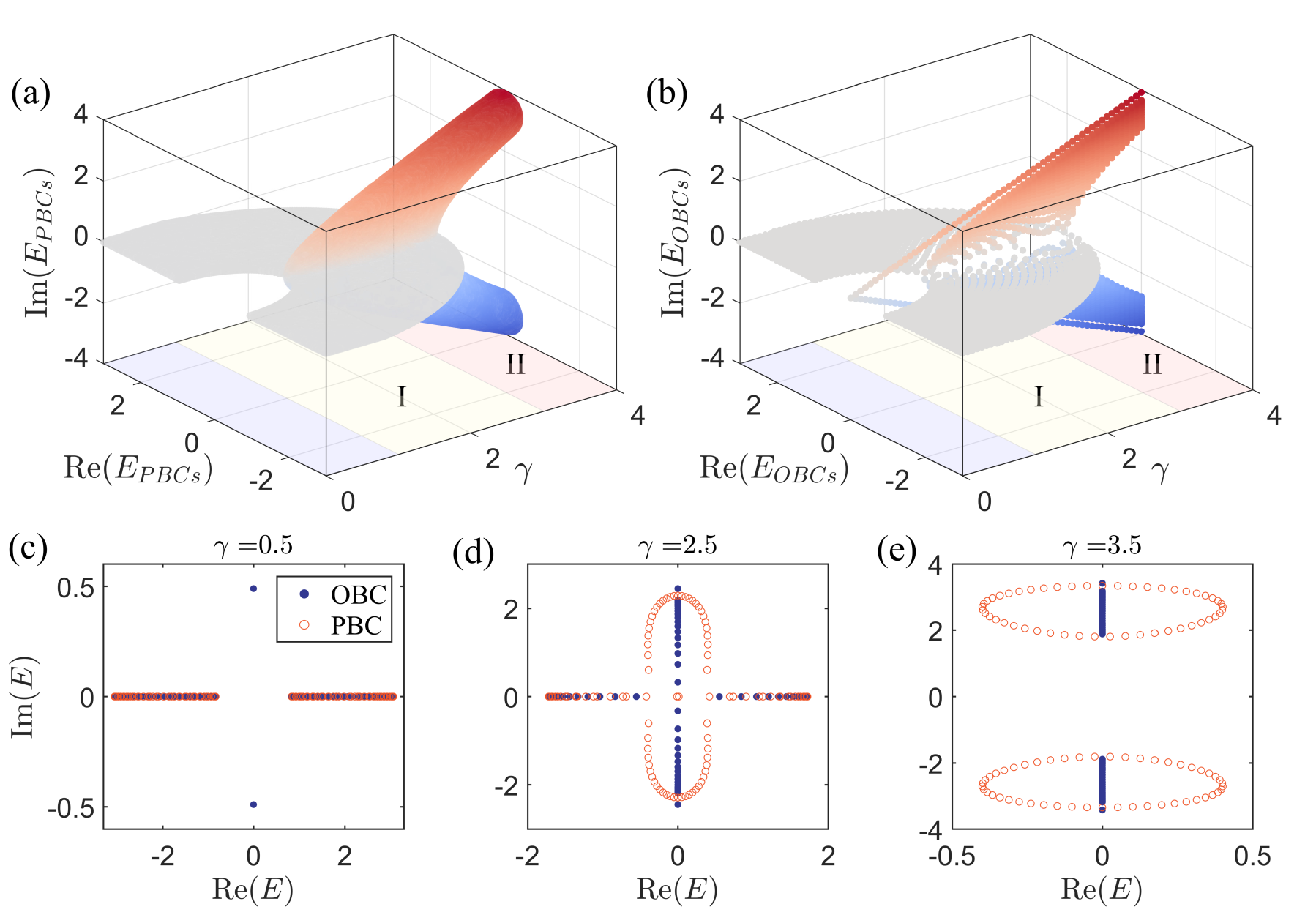}
\caption{Spectrum with the change of gain/loss under the periodic (a) and open (b) boundary conditions.
 Blue dots (orange circle) represent the energy under OBCs (PBCs) for (c) $\gamma=0.5$, (d) $\gamma=2.5$, and (e) $\gamma=3.5$, respectively. Here, the parameters are $t_{1}=1$, $t_{2}=2$, and $t_{3}=0.2$.
}%
\label{fig:2}%
\end{figure}

The energy of $h(k)$ is $E_{\pm}(k)=-2t_{3}\sin(k)\pm\sqrt{t_{1}^{2}+t_{2}^{2}-\gamma^{2}+2t_{1}t_{2}\cos(k)}$.
As shown in FIG. \ref{fig:2}, there are three different phases in our system. (1) $\mathcal{PT}$-symmetric phase: for $\gamma<|t_{1}-t_{2}|$, all eigenvalues of $h(k)$ are real, and there is no NHSE; (2) $\mathcal{PT}$-broken phase I: for $|t_{1}-t_{2}|<\gamma<|t_{1}+t_{2}|$, eigenvalues of $h(k)$ become complex. Due to the winding number
\begin{equation}
\mathcal{V}(E_b)=\frac{1}{2\pi i}\int_0^{2\pi}\frac{d\ln[\det(h(k)-E_b)]}{dk} dk
\end{equation}
is nonzero, NHSE emerges; (3) $\mathcal{PT}$-broken phase II: for $\gamma>|t_{1}+t_{2}|$, energy spectrum under PBC and open boundary conditions (OBCs) are two isolated loops and lines on complex energy plane, respectively. This also means the emergence of NHSE.

The NHSE means that all the eigenstates of an open chain are localized near the boundary. To describe this behavior, the non-Bloch band theory is introduced by the eigen-equation
\begin{align}
f(\beta,E) & = \mathrm{Det}[H(\beta)-E]  = \sum_{n=0}^{4} r_{n} \beta^{n}=0,
\end{align}
where solutions of $f(\beta,E)=0$ are labeled as $|\beta_{1}(E)|\le |\beta_{2}(E)|\le |\beta_{3}(E)|\le |\beta_{4}(E)|$ and the GBZ is given by the trajectory of $\beta_2$ and $\beta_3$ under the condition  $|\beta_{2}|=|\beta_{3}|$. With the $\mathcal{PT}$ symmetry and the anomalous particle-hole symmetry, we arrive at
\begin{equation}
f(\beta,E)=f(\frac{1}{\beta^{*}},E^*),\,\,\,\,f(\beta,E)=f(\beta^*,-E^*),
\end{equation}
which ensures that the GBZ appear in ($\beta,\frac{1}{\beta},\beta^*,\frac{1}{\beta^{*}}$) pairs. The GBZ implies two conclusions in our model. First, the states corresponding to real energy under OBCs are extended. Second, the states with Im$(E)$$>0$ and Im$(E)$$<0$ are localized at opposite ends.

Given that the hopping parameters $t_1$, $t_2$, and $t_3$ lack non-Hermitian properties, the complex spectrum is solely dependent on the gain and loss parameter $\gamma$. Therefore FIG. \ref{fig:2} shows that the OBC spectrum is either real or purely imaginary irrespective of the parameters $t_1$, $t_2$, $t_3$, and $\gamma$, which means the energy of the 1D model can only be either real or purely imaginary, corresponding to the extended states or skin states respectively under OBC.

\begin{figure}[hbth]
\includegraphics[width=0.5\textwidth]{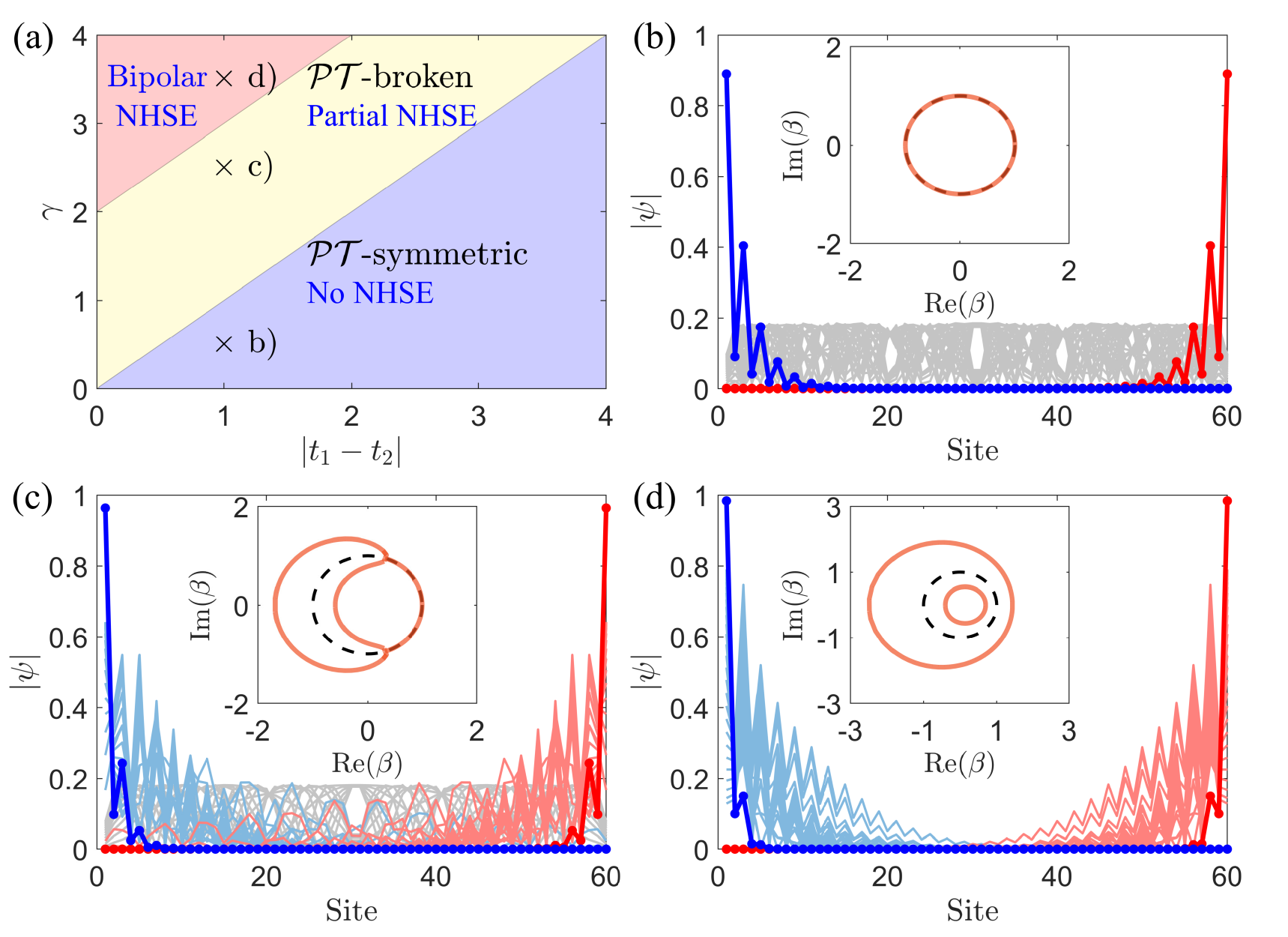}
\caption{(a) Phase diagram as a function of $t_1$ and $\gamma$ for $t_2=2$. (b)-(d) Spatial distributions of the eigenstates and corresponding GBZ. We use $\gamma=0.5$ in (b), $\gamma=2.5$ in (c), and $\gamma=3.5$ in (d). The extended states are represented by gray lines, while the eigenstates corresponding to eigenvalues with Im$(E)< 0$ (Im$(E)>0$) are shown with light blue (light red) lines.
The eigenstates with the largest imaginary parts of the eigenvalues are shown as dark blue (dark red) dashed lines. Other parameters are $t_1=1$, $t_2=2$, $t_3=0.2$, and $N = 30$.}%
\label{fig:3}%
\end{figure}

We present the phase diagram of our model in FIG. \ref{fig:3}. For the $\mathcal{PT}$-symmetric phase with $\gamma<|t_1-t_2|$,  as shown in FIG. \ref{fig:3}(a), the systems possess real energy spectra [see FIG. \ref{fig:2}(c)] and $\mathcal{PT}$-symmetric eigenfunctions. We can see from FIG. \ref{fig:3}(b) that the GBZ is as same as conventional Brillouin zone (BZ) and the absence of NHSE in the $\mathcal{PT}$-symmetric phase. For the $\mathcal{PT}$-broken phase I with $| t_1-t_2|<\gamma<|t_1+t_2|$, some eigenvalues become complex [see FIG. \ref{fig:2}(d)], and the spatial distribution of all eigenstates demonstrates the coexistence of extended states and NHSE, refer to partial NHSE, as shown in FIG. \ref{fig:3}(c). The corresponding GBZ is made up of three parts:  $|\beta|>1$,  $|\beta|=1$ and  $|\beta|<1$. In the regime of $|\beta|<1$ ($|\beta|>1$), eigenstates localize at the left (right) boundary, which indicate the direction of the NHSE. We can also determine the direction of the NHSE by the sign of the imaginary part of the OBC energy spectra. The eigenstates localize at the right (left) boundary for Im$(E)> 0$ (Im$(E)<0$). For the $\mathcal{PT}$-broken phase II, the PBC energy spectra become two circles on complex energy plane [see FIG. \ref{fig:2}(e)]. Correspondingly, the GBZ is composed of two circles with $|\beta|>1$ and $|\beta|<1$, respectively. The system exhibit a bipolar NHSE, which is similar to the $Z_2$ NHSE in spinful systems \cite{PhysRevResearch.2.043167,PhysRevLett.130.203605}. The emergence of such NHSE showing localized states at both ends of the systems can be predicted by a topological invariant
\begin{equation}
\mathcal{W}=\frac{\mathcal{V}^{+}(E_{b,+})-\mathcal{V}^{-}(E_{b,-})}{2} \,\,\, \mathrm{mod} \,\,\,\,2
\end{equation}
where $\mathcal{V}^{\pm}(E_{b,\pm})$ is the spectral winding number for the bands with Im($E$)$>$0 and Im($E$)$<$0, respectively. Our results show that we may realize tunable NHSE by tuning gain and loss ($\pm i\gamma$).

\textit{NHSEs in 2D non-Hermitian system}.---We can generalize the results of 1D non-Hermitian system to the corresponding 2D system [see FIG. \ref{fig:1}(b)] whose Hamiltonian is
\begin{equation}
H_{2D}=H_x\otimes \mathbb{I}+\mathbb{I}\otimes H_y,
\label{eq:3}
\end{equation}
\begin{figure}[hbth]
\includegraphics[width=0.5\textwidth]{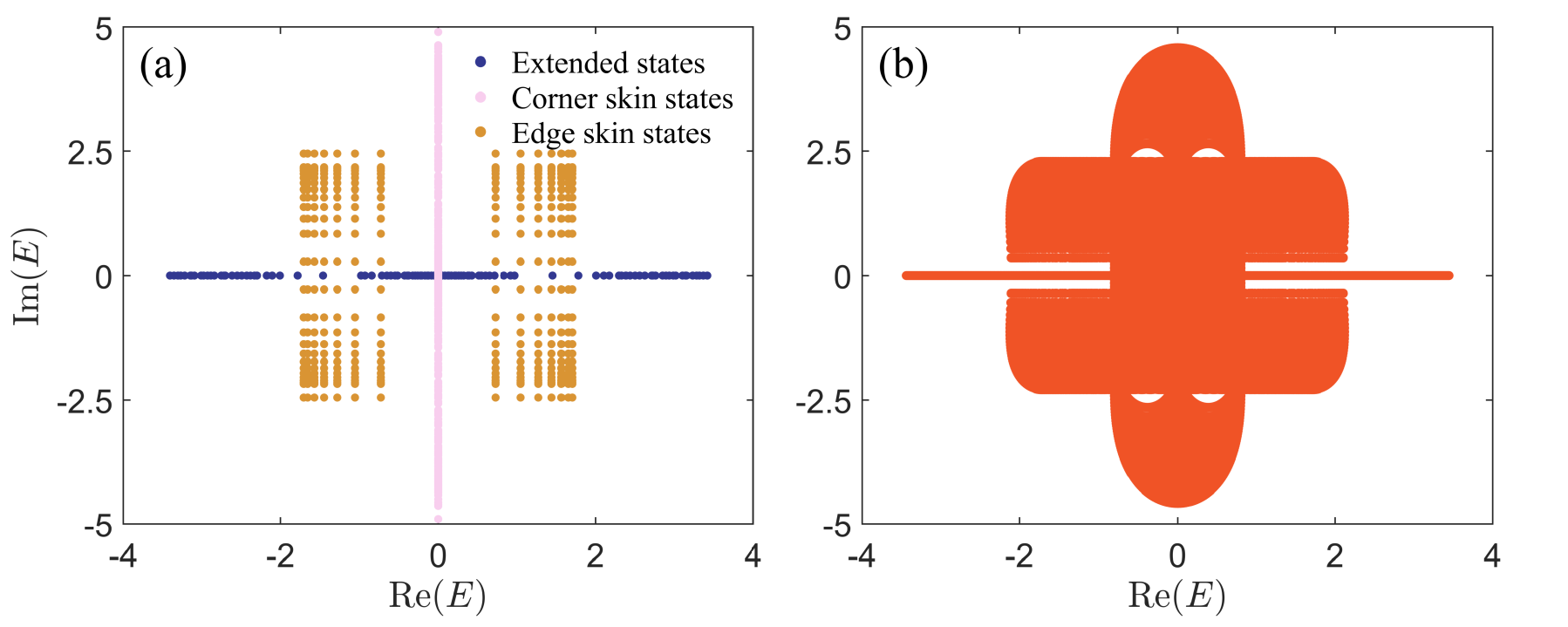}
\caption{The complex spectrum of the 2D non-Hermitian model under (a) OBCs and (b) PBCs. The parameters are given as $t_{1,x}=t_{1,y}=1$, $t_{2,x}=t_{2,y}=2$, $t_{3,x}=t_{3,y}=0.2$, $\gamma_x=\gamma_y=2.5$, and the system consists of 40×40 sites.
}%
\label{fig:4}%
\end{figure}
where $H_{\alpha}$ ($\alpha=x, y$) is the 1D Hamiltonian in Eq.\eqref{eq:1} with $(t_{1},t_{2},t_{3},\gamma)=(t_{1,\alpha},t_{2,\alpha},t_{3,\alpha},\gamma_{\alpha})$. The eigenvalue and corresponding eigenstate are $E=E_x+E_y$ and $\lvert \psi \rangle=\lvert \psi_x \rangle \otimes \lvert \psi_y \rangle$, where $H_\alpha\lvert \psi_{\alpha}\rangle=E_{\alpha}\lvert \psi_{\alpha}\rangle$.
In the 2D model, we may expect richer phenomena than in the 1D system, possibly including new types of NHSE.

We numerically obtain the complex spectrum of the 2D non-Hermitian model under OBCs and PBCs, as shown in FIG. \ref{fig:4}. The OBC spectrum, as shown in FIG. \ref{fig:4}(a), is composed of three parts --- the real spectrum, the purely imaginary spectrum, and the complex spectrum with nonzero real part and imaginary part, corresponding to blue, pink, and yellow dots, respectively. The 2D Hamiltonian corresponds to the appropriate 1D Hamiltonians $H_x$ and $H_y$ which exhibit partial NHSE and their eigenvalues must be either real or purely imaginary. The real 2D energy $E$ corresponds to the real 1D energies $E_x$ and $E_y$, both featuring extended states and free from NHSE; The purely imaginary 2D energy $E$ corresponds to the purely imaginary 1D energy $E_x$ and $E_y$, both featuring skin states and unavoidably leads to corner skin modes; The complex 2D energy $E$ corresponds to a real 1D energy $E_x$ plus a purely imaginary $E_y$ or vice versa, one with extended states and the other with skin states, naturally leads to edge skin modes. On the other hand, the PBC spectrum shown in FIG. \ref{fig:4}(b) is composed of two parts --- spectral arc (line) and spectral area. According to the theorem of universal bulk-boundary correspondence in two and higher dimensional non-Hermitian bands \cite{YZS2022}, the NHSE appears on generic open-boundary geometries if and only if the spectral area is nonzero. Therefore, we can establish the correspondence between 2D OBC spectrum and 2D PBC spectrum. There is no NHSE for the real part of the OBC spectrum, which matches exactly with the spectral line of the PBC spectrum. The NHSE is present in both the imaginary and complex parts of the OBC spectrum, matching the spectral region of the PBC spectrum.

\begin{figure*}[hbth]
\includegraphics[width=1\textwidth]{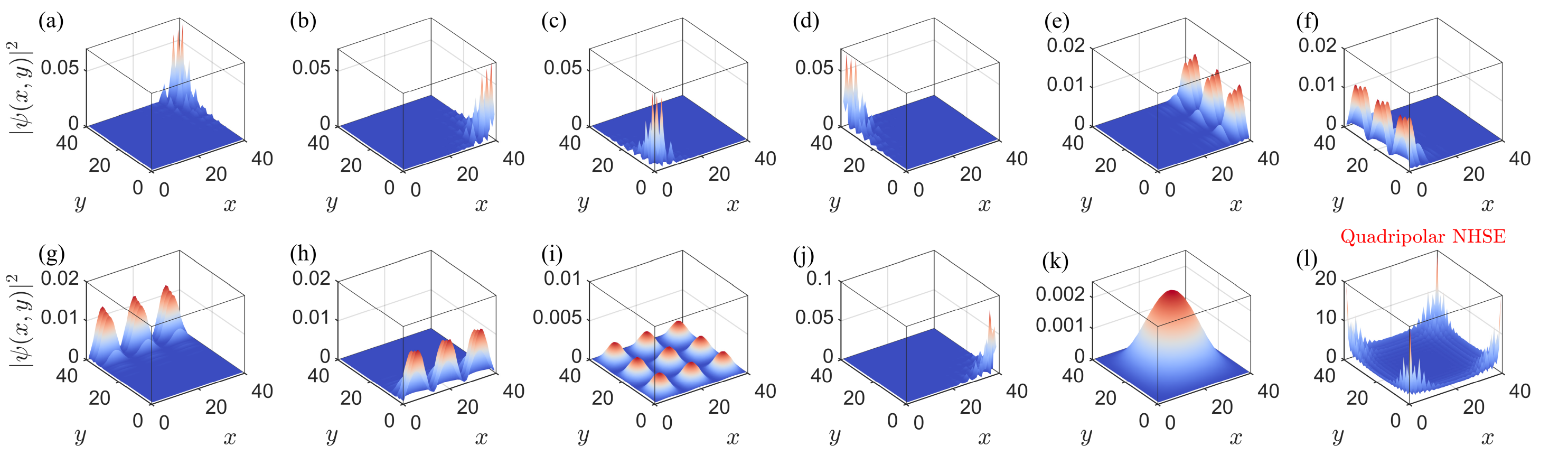}
\caption{The probability distributions of different eigenstates in a finite system consisting of 40×40 sites. (a-d) Corner skin modes which have eigenvalues (a) $E_x=1.58i$, $E_y=0.84i$, (b) $E_x=1.58i$, $E_y=-0.84i$, (c) $E_x=-1.58i$, $E_y=-0.84i$, (d) $E_x=-1.58i$, $E_y=0.84i$, respectively. (e-h) Edge skin states that have eigenvalues (e) $E_x=1.58i$, $E_y=1.57$, (f) $E_x=-1.58i$, $E_y=1.57$, (g) $E_x=1.57$, $E_y=1.58i$, (h) $E_x=1.57$, $E_y=-1.58i$, respectively. (i) Extended states with eigenvalues $E_x=E_y=1.57$. (j) Corner skin states with $E_x=-E_y=1.39i$. (k) Extended states with $E_x=-E_y=1.71$. (l) Probability distributions of sum of all eigenstates. We use $t_{1,x}=t_{1,y}=1$, $t_{2,x}=t_{2,y}=2$, $t_{3,x}=t_{3,y}=0.2$, and $\gamma_x=\gamma_y=2.5$.%Hybrid states with $E=0$ exhibit coexistence of extended states and corner skin states.
}%
\label{fig:5}%
\end{figure*}

For instance, if we take $E_x = 1.58i$ and $E_y = 0.84i$, then $E$ is purely imaginary. When $E_x = 1.58i$, the corresponding eigenstate is localized at the boundary ($L_x=40$) in the $x$-direction. Similarly, when $E_y = 0.84i$, the corresponding eigenstate is localized at the boundary ($L_y=40$) in the $y$-direction. Therefore, the eigenstates of $H_{2D}$ corresponding to the purely imaginary energy should be the corner skin modes [see FIG. \ref{fig:5}(a)-(d)].
For complex eigenvalues $E$ with both real and imaginary parts, the states are localized in an edge [FIG. \ref{fig:5}(e)-(h)]. All states with real eigenvalues $E$ are extended [see FIG. \ref{fig:5}(i)]. In this system, we can obtain various skin states and extended states, which is useful for exploring their applications. Such a unique feature has yet to be found in non-Hermitian systems with gain and loss. For the 2D OBC system, as shown in FIG. 5(j) and FIG. 5(k), the zero-energy states result from states with opposite real or imaginary energies, corresponding to hybrid states composed of corner skin states and extended states. Due to the probability distributions of sum of all eigenstates being mainly localized at the four corners, this phenomenon is called the quadripolar NHSE [see FIG. \ref{fig:5}(l)].

\textit{Experimental proposals}.---Recent years, electric circuits provide an excellent platform to study non-Hermitian topological phases which is difficult to realize in  quantum systems \cite{PhysRevLett.114.173902,PhysRevB.100.165419,Wang2020,Yu2020,PhysRevResearch.2.023265,PhysRevResearch.2.022062,PhysRevLett.126.146802,PhysRevB.105.195127,PhysRevB.107.085426,nanolett2022,PhysRevResearch.5.043034,Kim2023,PhysRevB.107.184108,PhysRevApplied.20.064042,Zhu2023,Zhang2024,PhysRevApplied.21.054013,PhysRevApplied.21.034043,PhysRevB.109.115407}. Here, we propose a scheme to show our results in electric circuit system.

\begin{figure}[hbth]
\includegraphics[width=0.47\textwidth]{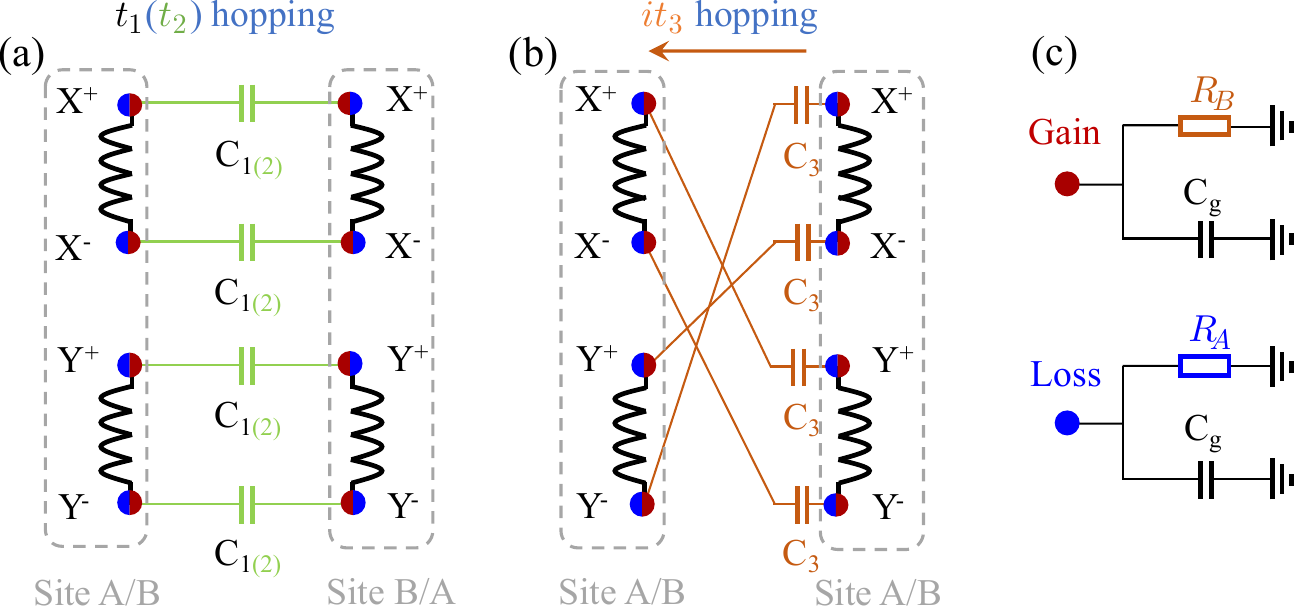}
\caption{Structure of the coupling elements between lattice sites. (a-b) show the linear circuit connections that is designed to simulate the nearest-neighbor and next-nearest-neighbor interactions, respectively. (c) The simulation gain and loss by $R_B=-R$ ($R_A=R$).
}%
\label{fig:6}%
\end{figure}

The electric circuit corresponding our model Eq.(\ref{eq:1}) is shown in FIG. \ref{fig:6}, which consists of capacitors, inductors, and resistors. The hoppings $t_1$, $t_2$, and $t_3$ are represented by capacitors $C_1$, $C_2$ and $C_3$, respectively. And the on-site gain (loss) corresponds to resistive elements $-R$ ($R$) grounded. Each lattice site is equipped with two inductors $X$ and $Y$ whose ends are labeled as $X^{\pm}$ and $Y^{\pm}$, and the voltages across the inductors denoted as $U_{X(Y)}=V_{X(Y)^{+}}-V_{X(Y)^{-}}$. All inductors have the same inductance $L$. For next-nearest-neighbour hoppings, a phase of $\pi/2$ is induced by braiding the capacitive couplings \cite{PhysRevX.5.021031,PhysRevB.99.115410,Yang2021,PhysRevLett.126.146802,Zhang2023,PhysRevLett.130.206401}.

Here, the relation between the currents and voltages is given by Kirchhoff’s law
\begin{eqnarray}
\boldsymbol{I}(\omega)=J(\omega) \boldsymbol{V}(\omega),\label{eq:dianlu}
\end{eqnarray}
where $\boldsymbol{V}$ and $\boldsymbol{I}$ are voltage and current entering from the external source, and $\omega$ is the AC driving frequency of the system. For convenience, the grounding capacitance is set as $C_{g}=C$, and capacitors corresponding to hoppings are set as $C_{1}=t_1C$, $C_{2}=t_2C$, and $C_{3}=t_3C$, where $C$ acts as a reference capacitance. Each pair of the $LC$ circuit has the same resonance frequency $\omega_0=1/\sqrt{LC}$. The voltages across the inductors are defined as $U_{n,X}^{A(B)}=V_{n,X^+}^{A(B)}-V_{n,X^-}^{A(B)}$. Considering current conservation, namely, the vanishing sum of the inflow and outflow currents at every node, Eq. \eqref{eq:dianlu} can be written as
\begin{eqnarray}
\begin{aligned}
U_{n,X}^{A} &= -\frac{\omega^2}{2\omega_0^2} \Big[
-\left (1+\frac{1}{iR_A\omega C}+t_1+t_2+ 2t_3  \right ) U_{n,X}^{A}\\
&+t_1 U_{n,X}^{B}+t_2 U_{n-1,X}^{B}-t_3 U_{n-1,Y}^{A}+t_3 U_{n+1,Y}^{A}\Big ],\\
U_{n,Y}^{A} & = -\frac{\omega^2}{2\omega_0^2} \Big[
-\left (1+\frac{1}{iR_A\omega C}+t_1+t_2+ 2t_3  \right ) U_{n,Y}^{A}\\
&+t_1 U_{n,Y}^{B}+t_2 U_{n-1,Y}^{B}+t_3 U_{n-1,X}^{A}-t_3 U_{n+1,X}^{A}\Big ].
\end{aligned}
\end{eqnarray}
We can also derive the equations for the inductor $X(Y)$ at the site ($n,B$):
\begin{eqnarray}
\begin{aligned}
U_{n,X}^{B} &= -\frac{\omega^2}{2\omega_0^2} \Big[
-\left (1+\frac{1}{iR_B\omega C}+t_1+t_2+ 2t_3  \right ) U_{n,X}^{B}\\
&+t_1 U_{n,X}^{A}+t_2 U_{n+1,X}^{A}-t_3 U_{n-1,Y}^{B}+t_3 U_{n+1,Y}^{B}\Big ],\\
U_{n,Y}^{B} &= -\frac{\omega^2}{2\omega_0^2} \Big[
-\left (1+\frac{1}{iR_B\omega C}+t_1+t_2+ 2t_3  \right ) U_{n,Y}^{B}\\
&+t_1 U_{n,Y}^{A}+t_2 U_{n+1,Y}^{A}+t_3 U_{n-1,X}^{B}-t_3 U_{n+1,X}^{B}\Big ].
\end{aligned}
\end{eqnarray}
The above equations can be expressed into an eigen-equation. After performing the Fourier transformation, we can obtain
\begin{eqnarray}
\Omega\begin{pmatrix}
 U_{k,\uparrow }^A\\
 U_{k,\uparrow }^B
\end{pmatrix} & = &
\begin{pmatrix}
p_{k}+i \gamma_{A}&    T_{k}\\
   T_{k}^{*}      & p_{k}+i \gamma_{B}
\end{pmatrix}
\begin{pmatrix}
 U_{k,\uparrow }^A\\
 U_{k,\uparrow }^B
\end{pmatrix},
\end{eqnarray}
where $U_{\uparrow ,\downarrow }=U_{X}\pm iU_{Y}$, $\Omega=\left ( t_1+t_2+ 2t_3 +1-\frac{2\omega_0^2}{\omega^2} \right )$. With the basis of $\left(U_{k,\uparrow}^A,U_{k,\uparrow}^B\right)^{T}$, $T_{k}=t_1+t_2e^{-ik}$ and $p_{k}= it_3e^{ik}-it_3e^{-ik}=-2t_3 \sin{k} $, and the non-Hermitian term is expressed as
\begin{eqnarray}
\gamma_{A(B)} & = & \frac{1}{R_{A(B)}}\sqrt{\frac{L}{C}},\,\,\,\,\, R_A+R_B=0.
\end{eqnarray}
It is important to note that $-R$ can be utilized using negative impedance converters with current
inversion (INICs) \cite{Imhof2018,PhysRevLett.126.215302}, corresponding to gain in the non-Hermitian SSH model. With the above proposal, we can carry out a feasible electrical-circuit simulation for our model.

If we generalize 1D circuits to 2D circuits, the above arguments remain valid, since the hoppings in $x$ and $y$ directions are the same as those in 1D systems. The only difference is that we need to redefine gain/loss elements in the $x$ and $y$ directions. The corresponding non-Hermitian terms are
\begin{eqnarray}
\begin{aligned}
\bar{\gamma_{A}} & = -(\frac{1}{R_{x}}+\frac{1}{R_{y}})\sqrt{\frac{L}{C}},\,\,\,\,\,\bar{\gamma_{B}}= -(\frac{1}{R_{x}}-\frac{1}{R_{y}})\sqrt{\frac{L}{C}},\\
\bar{\gamma_{C}} & = (\frac{1}{R_{x}}-\frac{1}{R_{y}})\sqrt{\frac{L}{C}},\,\,\,\,\,\,\,\,\bar{\gamma_{D}}= (\frac{1}{R_{x}}+\frac{1}{R_{y}})\sqrt{\frac{L}{C}},
\end{aligned}
\end{eqnarray}
where $R_{x}$ and $R_{y}$ are the grounding resistors in the $x$ and $y$ directions, respectively.

\textit{Summary}.---In summary, we have investigated tunable NHSE in systems with gain and loss. For 1D case, we find three phases---the $\mathcal{PT}$-symmetric phase, the $\mathcal{PT}$-broken phase I with partial NHSE, and the $\mathcal{PT}$-broken phase II with bipolar NHSE. The generalization to the 2D system allows us to unveil a quadripolar NHSE with localized states at four corners of the systems. Our proposal introduces several NHSE within a single system and significantly enhances the controllability of the non-Hermitian systems without resort to non-reciprocal hoppings. We also show a experimental realization of our model in the electric circuit systems.

This work has been supported by the research foundation of Institute for Advanced Sciences of CQUPT (Grant No. E011A2022328).
\bibliographystyle{apsrev4-1}
\bibliography{Reference}

\end{document}